# Design of a waveguide-coupled GeSn disk laser

Bahareh Marzban
*Institute of Integrated Photonics*
*RWTH Aachen University*
Aachen, Germany
fmarzban@iph.rwth-aachen.de

Jovana Nojić
*Institute of Integrated Photonics*
*RWTH Aachen University*
Aachen, Germany
jnojic@iph.rwth-aachen.de

Daniela Stange
*Forschungszentrum Jülich*
*Peter Grünberg Institute*
Jülich, Germany
https://orcid.org/0000-0002-2545-5832

Dan Buca
*Forschungszentrum Jülich*
*Peter Grünberg Institute*
Jülich, Germany
https://orcid.org/0000-0002-3692-5596

Jeremy Witzens
*Institute of Integrated Photonics*
*RWTH Aachen University*
Aachen, Germany
https://orcid.org/0000-0002-2896-7243

*Abstract*—We report on the design of a waveguide coupled GeSn microdisk-laser cavity in which the germanium virtual substrate serving as a template for GeSn growth is repurposed for the definition of passive on-chip interconnection waveguides. A main challenge resides in transferring the optical power from the upper (Si)GeSn gain stack to the underlying virtual substrate layer and is solved with laser mode engineering. Designs are based on experimentally realized layer stacks and waveguide outcoupling efficiencies as high as 27% are shown in compact resonator geometries with a small, 7 μm radius, with 42% of the power being recycled in the laser cavity.

## I. Introduction

Since its first demonstration as a direct bandgap semiconductor [1], germanium-tin (GeSn) has become a promising material for realizing CMOS-compatible laser sources for the silicon photonics (SiP) platform. Several groups have reported lasing from optically pumped GeSn lasers [2],[3], with recent demonstration of lasing at temperatures up to 273 K [4]. While under-etched structures allow relaxation of compressive strain and improvement of the directness of the material [5], non-under-etched cavities [6] facilitate the outcoupling of light to on-chip waveguides and may facilitate heat-sinking of electrically pumped structures in the future. GeSn layers are typically grown on a germanium virtual substrate (Ge-VS), which can be used for defining interconnection waveguides and coupling the light out of the laser cavity. In this work, we propose a novel coupling scheme for an electrically pumped GeSn laser, that features waveguides fabricated in the Ge-VS.

## II. Cavity and Waveguide Design

The laser is based on a vertical 700 nm GeSn/SiGeSn stack grown on a Ge-VS with compositions close to the ones used for an optically pumped laser in [5]. p- and n-doping are assumed to be added to the layer stack in order to enable electrical connectivity. It is crucial to grow the GeSn stack on a high-quality Ge-VS with low surface roughness, that can be obtained for a thickness above 1 μm [7]. Consequently, we choose a Ge-VS thickness of 1.2 μm for our design. The layer stack itself consists of (i) a 250 nm p-doped GeSn buffer layer with a step-graded Sn concentration (150 nm $GeSn_{0.08}$ doped $2e18$ $cm^{-3}$ followed by 100 nm $GeSn_{0.1}$ doped $1e18$ $cm^{-3}$), (ii) a 15 nm, highly p-doped ($1e19$ $cm^{-3}$) $Si_{0.08}GeSn_{0.1}$ layer, (iii) 6x 25 nm wide $GeSn_{0.135}$ quantum wells separated by 20 nm $Si_{0.05}GeSn_{0.12}$ barrier layers, topped by (iv) a 15 nm highly n-doped ($1e19$ $cm^{-3}$) $Si_{0.08}GeSn_{0.1}$ layer and (v) a 170 nm step-graded n-doped GeSn capping layer (100 nm $GeSn_{0.1}$ doped $2e18$ $cm^{-3}$ followed by 70 nm $GeSn_{0.08}$ doped $3e18$ $cm^{-3}$).

The step-graded GeSn buffer is used to confine the misfit dislocations at the interface to the Ge-VS [5] and between the buffer layers [4], keeping them away from the active region. The GeSn buffer (i) and capping (v) layers, as well as the highly doped $Si_{0.08}Ge_{0.82}Sn_{0.1}$ layers [(ii) and (iv)], also serve for lateral current transport from electrodes placed inside the disk [Fig. 1(a)] to the active region at the disk's periphery. Energy barriers for holes and electrons, respectively at the heterojunctions between the GeSn buffer (i) and the first SiGeSn barrier layer (ii) and between the GeSn capping layer (v) and the topmost SiGeSn barrier layer (iv), are lowered by the high doping of these SiGeSn layers. The active layer stack in-between is left undoped, but assumed to be intrinsically p-doped (~ $2e17$ $cm^{-3}$) due to native point defects [8].

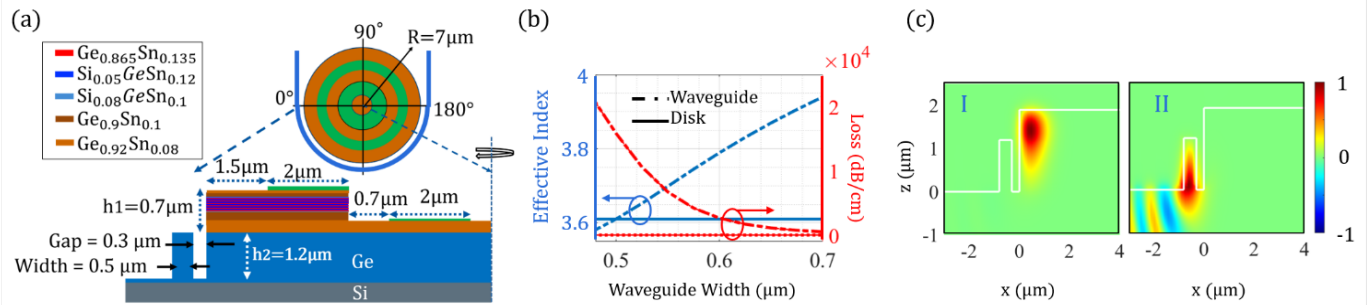

Fig. 1. (a) Microdisk laser with a bus waveguide separated by a 300 nm gap. (b) Effective index and losses of bus waveguide ground mode (TE0) as a function of its width. (c) Electric field of (I) the TE0 whispering gallery mode of the laser cavity and (II) the TE0 waveguide mode with a 0.5 μm waveguide width.

As in other electrically pumped microdisk lasers, light remains close to the outer edge of the disk in the form of whispering gallery modes. Placement of electrodes inside of the disk thus reduces optical absorption losses associated to metal electrodes and contacting materials such as nickel-stanogermanides (NiGeSn), but may lead to large excess currents given the high dark recombination rates observed in current gain materials. Since GeSn layers experience compressive strain, the laser is expected to lase in the TE polarization.

From modeling [9] of the optically pumped multi-quantum-well GeSn/SiGeSn microdisk laser with experimental results published in [5], we estimate the electron concentration at threshold to be 3.5e17 cm$^{-3}$ inside the wells at 20 K and the modal gain to be 600 cm$^{-1}$, pointing to high internal losses due to surface roughness as well as inter-valence-band and inter-conduction-valley free carrier absorption. Optical losses in an electrically pumped structure will be even higher due to the doping required for carrier transport. Consequently, in long laser cavities, most of the generated light will be reabsorbed prior to reaching output waveguides. To maintain a reasonable wall-plug efficiency, we choose a relatively small, 7 μm radius. In the following, we will estimate the ratio between the power outcoupled to the interconnect waveguide ($P_{out}$) and the power generated by the laser ($P_{gen}$) as

$$\frac{P_{out}}{P_{gen}} = \frac{T_{WG}}{1-T} \frac{\frac{1}{L}\ln\left(\frac{1}{T}\right)}{\alpha + \frac{1}{L}\ln\left(\frac{1}{T}\right)} \quad (1)$$

with $T_{WG}$ the (power-)coupling coefficient from the disk mode to the waveguide, $T$ the power remaining in the disk mode after crossing the waveguide coupling section, $L$ the circumference of the disk, and $\alpha$ the loss coefficient of the disk (in inverse length). The efficiency could be even further increased by further shrinking the structure. We have, however, opted for the 7 μm radius for the electrical contacting of the structure to remain practicable.

Interconnect waveguides are assumed to be defined in the Ge-VS, after removal of the SiGeSn/GeSn stack, by etching down to the silicon. Here, high coupling strengths are required to get reasonable power extraction efficiencies, due to the high internal laser losses [Eq. (1)]. The most commonly used outcoupling method for microdisks is evanescent coupling of the cavity mode to a waveguide. If a straight waveguide is used, the low junction length will result in a low coupling strength. To achieve high coupling strengths, one could wrap a curved waveguide around the laser [Fig. 1(a)].

In our case, this method proved very difficult to implement. The main difficulty arises from the cavity mode being located at a different vertical position than the waveguide mode. To solve the vertical mismatch, one could use a racetrack cavity instead of a circular cavity and place a mode converter in the straight section of the racetrack. However, in order to have an adiabatic transition from the active region to the Ge waveguide, a mode converter with a transition length of several tens of μm would be necessary [10]. Since the narrow portions of the tapered transitions would be very difficult to contact in a non-buried structure, the active material would not be uniformly pumped, which would in turn result in extremely high losses. Thus, we will first analyze coupling from the laser to a wrapped waveguide placed 300 nm away from the cavity without an intra-cavity mode converter. After explaining the challenges associated to such a design, we will present our proposed solution.

The TE ground mode of the circular cavity has the highest confinement factor among the modes present in the microdisk laser and is equal to 0.18. Therefore, we consider the coupling of a curved waveguide to the TE0 mode of the disk, which has an effective index equal to 3.61. Given the 7 μm radius of the microdisk, this mode has a bending loss of 0.98 dB/cm, which is negligible compared to the intrinsic material loss. The minimum waveguide width which supports a TE mode is equal to 0.48 μm. The effective indices and bending losses of the waveguide's TE0 mode as a function of waveguide width can be seen in Fig. 1(b). To achieve high coupling strengths, the phase accumulated by the two modes should remain equal in the coupling section, i.e., the two modes need to be phase matched. To achieve this condition $n_{WG} \cdot r_{WG} = n_{MD} \cdot r_{MD}$ should be satisfied, where $n_{WG}$, $n_{MD}$ are the effective indices of the waveguide and microdisk, and $r_{WG}$, $r_{MD}$ their radii. This is achieved for a waveguide width of about 0.5 μm assuming a 300 nm gap to the microdisk. However, due to the low index contrast at the Si-Ge interface at the bottom of the waveguide core, bending losses of the curved waveguide at the chosen radius become extremely high as the waveguide width is reduced.

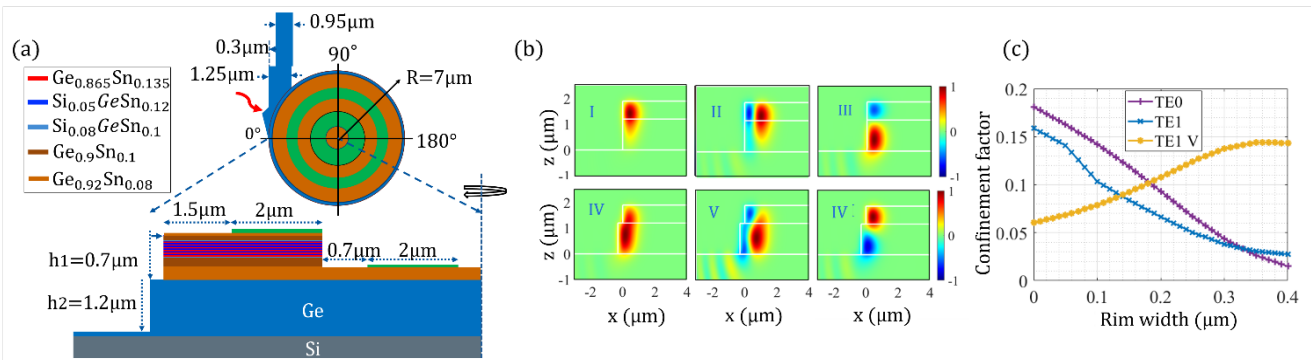

Fig. 2. (a) Proposed laser cavity with dimensions. (b) Electric field of whispering gallery (I) TE0, (II) TE1, (III) TE1V modes of the structure without rim (top panels) and of (IV) TE0, (V) TE1, (VI) TE1V modes of the structure with a 350 nm rim (bottom panels). (c) Confinement factors vs. rim width for the three TE-modes TE0, TE1, and TE1V.

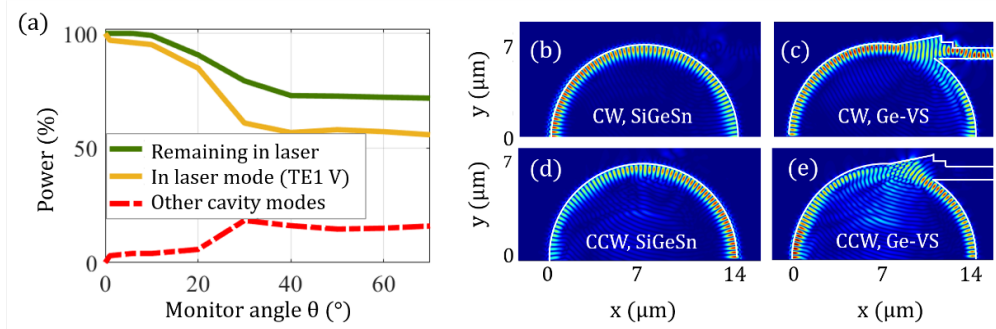

Fig. 3. (a) Evolution of power circulating inside the disk as a function of azimuthal angle, while crossing the waveguide junction. Field of the (b), (c) CW- and (d), (e) CCW-modes in the center-planes of the SiGeSn stack [(b): CW, (d): CCW] and of the Ge-VS [(c): CW, (e): CCW].

As the waveguide width is reduced to 0.5 μm, the effective index of the waveguide's TE0 mode approaches that of the laser mode, but the bending losses reach 15,000 dB/cm, which is clearly not acceptable. Moreover, even with the small gap of 300 nm, the two modes of the two structures do not couple to each other, as can be seen from the field profiles shown in Figs. 1(c), due to the low mutual overlap resulting from the vertical mismatch. Thus, it appears extremely challenging to obtain high coupling strengths with this method in this structure.

To circumvent these problems, we introduce a 350 nm wide step-like rim at the outer edge of the cavity, where the SiGeSn stack has been etched but not the underlying Ge-VS [Fig. 2(a)]. This significantly modifies the properties of the three dominant TE modes as shown in Figs. 2(b) and 2(c). Even higher order modes experience bending losses above 1000 dB/cm and do not play a role here. By introducing the rim, the TE0 (I, IV) and TE1 (II, V) modes are pulled down into the Ge-VS. This can be simply understood with conformal mapping, since, after transforming the effective index to the equivalent one in an equivalent, rectilinear geometry, the transformed index at the outer edge of the Ge-VS region is boosted as a consequence of the enlarged radius [11]. On the other hand, the TE mode with two vertical lobes (III, VI), TE1V, increases its overlap with the gain region. Thus, after introduction of the rim, the laser is expected to lase in the TE1V mode, that has a substantially higher confinement factor than the other two.

A Ge-VS waveguide with a width equal to 1.25 μm and an outer edge tangent to the circumference of the Ge-VS disk is added to the structure [Fig. 2(a)] and routes out the bottom lobe of the circulating TE1V mode. To direct it into the TE0 waveguide mode, a 2nd waveguide with a smaller width of 950 nm is appended to the 1st with an offset. With this scheme $T = 56\%$ of the power stays in the TE1V mode [Fig. 3(a)], while 23% is coupled to the waveguide. Additionally, 16% of the power is scattered to other cavity modes and 5% scattered to the substrate and to air at the junction. Of the 23% coupled to the waveguide, only 72% end up in the ground mode of the second waveguide, resulting in $T_{WG} = 16\%$. Reducing the rim width to 250 nm would result in improved results with $T = 42\%$ and $T_{WG} = 27\%$, however this would also reduce the contrast between confinement factors of competing modes, potentially compromising the laser selecting TE1V. Given the estimated $\alpha > 600$ cm$^{-1}$, Eq. (1) is evaluated at $< 6.5\%$ and $< 11.5\%$, respectively for a rim of 350 and 250 nm. It should be noted that some amount of excess losses is unavoidable in this outcoupling scheme, as only the upper lobe of the TE1V mode is transmitted across the microdisk junction, resulting in some mode mismatch. Nevertheless, the laser is primarily penalized by its high internal losses.

In order to suppress the counter clockwise (CCW) propagating mode, we added a small notch in the outcoupling section indicated by a red arrow in Fig. 2(a). It acts as a mirror and partially back-reflects it into the clockwise (CW) mode, breaking the symmetry and favoring the latter. Field profiles of the CW and CCW modes around the waveguide junction can be seen in Figs. 3(b)-3(e).

As an additional benefit resulting from the increased overlap of the hybrid TE1V mode with the low loss Ge-VS, higher output power lasers may be realized [12].

III. CONCLUSIONS

Design studies show that it is extremely challenging to achieve a substantial outcoupling efficiency from GeSn microdisks with a small radius to evanescently coupled waveguides defined in the Ge-VS. As an alternative coupling scheme not requiring adiabatic tapers, we design of microdisk configuration expected to lase into mixed modes with partial overlap in both the gain medium and the underlying Ge-VS, facilitating outcoupling of the light. This preprint is an extended version of a conference proceeding paper from the 2020 IEEE Summer Topicals.